\documentclass[]{aastex631}

\usepackage{amsmath}
\usepackage{amssymb}
\usepackage{enumerate}

\newcommand{\athena}{{Athena++}}

\newcommand{\arepo}{{Arepo}}

\newcommand{\be}{\begin{eqnarray}}
\newcommand{\ee}{\end{eqnarray}}

\newcommand{\csq}{\ensuremath{c_{s}^{2}}}
\newcommand{\mdot}{\ensuremath{\dot{M}}}

\newcommand{\vel}{\ensuremath{\boldsymbol{v}}}

\newcommand{\rsink}{\ensuremath{R_{\rm sink}}}
\newcommand{\rsonic}{\ensuremath{r_{\rm sonic}}}
\newcommand{\ra}{\ensuremath{R_{A}}}
\newcommand{\rs}{\ensuremath{R_{s}}}
\newcommand{\fdf}{\ensuremath{F_{\rm DF}}}

\newcommand{\mach}{\ensuremath{\mathcal{M}}}
\newcommand{\facc}{\ensuremath{F_{\rm acc}}}

\newcommand{\lp}[1]{\textbf{\textrm{\color{red} #1}}}

\renewcommand{\sout}[1]{}
\renewcommand{\lp}[1]{#1}

\begin{document}

\title{Morphology and Mach Number Dependence of Subsonic Bondi-Hoyle Accretion}

\correspondingauthor{Logan J. Prust}
\email{ljprust@kitp.ucsb.edu}

\author{Logan J. Prust}
\affiliation{Kavli Institute for Theoretical Physics, University of California, Santa Barbara, CA 93106, USA}

\author{Hila Glanz}
\affiliation{Technion -- Israel Institute of Technology, Haifa, 3200002, Israel}

\author{Lars Bildsten}
\affiliation{Kavli Institute for Theoretical Physics, University of California, Santa Barbara, CA 93106, USA}
\affiliation{Department of Physics, University of California, Santa Barbara, CA 93106, USA}

\author{Hagai B. Perets}
\affiliation{Technion -- Israel Institute of Technology, Haifa, 3200002, Israel}

\author{Friedrich K. Röpke}
\affiliation{Heidelberger Institut für Theoretische Studien, Schloss-Wolfsbrunnenweg 35, 69118 Heidelberg, Germany}
\affiliation{Zentrum für Astronomie der Universität Heidelberg \\
Institut für theoretische Astrophysik, Philosophenweg 12, 69120 Heidelberg, Germany}



\begin{abstract}

We carry out three-dimensional computations of the accretion rate onto an object (of size $R_{\rm sink}$ and mass $m$) as it moves through a uniform medium at a subsonic speed $v_{\infty}$. The object is treated as a fully-absorbing boundary (e.g. a black hole). In contrast to early conjectures, we show that when $R_{\rm sink}\ll R_{A}=2Gm/v^2$ the accretion rate is independent of $v_{\infty}$ and only depends on the entropy of the ambient medium, its adiabatic index, and $m$. Our numerical simulations are conducted using two different numerical schemes via the \athena\ and \arepo\ hydrodynamics solvers, which reach nearly identical steady-state solutions. We find that pressure gradients generated by the isentropic compression of the flow near the accretor are sufficient to suspend much of the surrounding gas in a near-hydrostatic equilibrium, just as predicted from the spherical Bondi-Hoyle calculation. Indeed, the accretion rates for steady flow match the Bondi-Hoyle rate, and are indicative of isentropic flow for subsonic motion where no shocks occur. We also find that the accretion drag may be predicted using the Safronov number, $\Theta=R_{A}/R_{\rm sink}$, and is much less than the dynamical friction for sufficiently small accretors ($R_{\rm sink}\ll R_{A}$).

\end{abstract}



\section{Introduction} \label{sec:intro}

Over many decades, the problem of Bondi-Hoyle-Lyttleton (BHL) accretion flow has been studied through analytical estimates and numerical calculations in both two and three dimensions, beginning with \citet{1971MNRAS.154..141H}. These studies have covered a range of adiabatic indices and have treated stationary accretors as well as those with either subsonic or supersonic velocities, aiming to determine the rates of mass and momentum accretion as well as the shock structure (if present). 


BHL flows are relevant to many astrophysical problems, particularly to the accretion onto black holes. Other accreting objects have hydrodynamically relevant surfaces which limit their accretion -- such as protoplanets within disks -- though BHL may still be used to obtain an upper bound for the accretion rate \citep{2003ApJ...589..578N}. Wind accretion onto stars and X-ray binaries is a commonly-used application of BHL flows, though this typically involves supersonic motion and is therefore not relevant to this work \citep[see, for example,][]{2004MNRAS.347..173S}. On the other hand, galaxies in clusters can move through the intergalactic medium (IGM) subsonically due to the high temperature of the IGM \citep{2004EdgarReview}. Thus subsonic BHL flow may describe accretion onto galaxies from the IGM, as observed by \citet{1980ApJ...242..511D}, though galactic outflows may complicate this process \citep{2000ApJ...543..611O}. Finally, binary interactions such as common envelope evolution involve the subsonic or supersonic motion of an accretor through a stellar envelope. Here the axisymmetry of BHL flow is broken by density gradients in the envelope, creating an angular momentum barrier to accretion. Thus the accretion rate must be modified according to the density scale height of the envelope \citep{1989ApJ...337..849T,2015ApJMacLeod}. These applications vary in terms of their velocities, accretion radii, and adiabatic indices ($\gamma$), necessitating a general formulation for the accretion rate.

Several attempts have been made to determine a self-similar solution for the accretion flow and to derive an analytical form for the accretion rate $\mdot$. However, they have struggled to provide generality in $\gamma$ and to bridge the gap between stationary, subsonic, and supersonic accretors. The solution for spherical flow of \citet{1952MNRAS.112..195B} is well-known, as is the accretion rate for zero-temperature axisymmetric flow \citep{1939PCPSHoyleLyttleton}, which is relevant to highly-supersonic velocities. For intermediate Mach numbers $0<\mach\lesssim2$, the Bondi rate is often interpolated such that it reproduces these results in the limits $\mach\rightarrow0$ and $\mach\rightarrow\infty$, respectively \citep{1952MNRAS.112..195B}. Additionally, \citet{ruffert3} obtained fitting formulae from the results of their numerical simulations, and subsequently \citet{1997A&A...320..342F} derived an analytical interpolation between subsonic and supersonic flows for $\gamma>9/7$. \lp{These predictions agree with their numerical results for $\mach=0.6$ and 1.4, and suggest that the accretion rate is equal to the spherical Bondi rate for at least some range of Mach numbers. However, the behavior of $\mdot$ over the full range of subsonic flow is not known.}

The relevant length scales are the accretion radius, defined in \cite{1939PCPSHoyleLyttleton} as
\be
\ra = \frac{2 G m}{v_{\infty}^{2}},
\ee
as well as the Bondi radius
\be
R_{\rm B} = \frac{2 G m}{c_{s,\infty}^{2}}.
\ee
Here $m$ is the mass of the accretor, $v_{\infty}$ is its velocity relative to the ambient medium, and $c_{s,\infty}$ is the sound speed. We primarily use $\ra$ throughout this work for consistency between our simulations and for comparison to previous works \citep{ruffert2, ruffert3, 2015ApJMacLeod}. The astronomical objects listed above vary in the ratio of their physical radii $R$ to $\ra$, ranging from $R/\ra\sim1$ for galaxies and planets down to $10^{-5}$ --  $10^{-10}$ for black holes. \lp{As} there is no \textit{a priori} reason to believe that the accretion flow is described by the same solution in both regimes, \lp{numerical calculations are needed to determine which accretion flows may be described by the BHL model.}

In this paper, we aim to characterize the flow around both large and small accretors relative to $\ra$, and in particular to determine the accretion rate. We organize this paper as follows. We review the relevant theory to determine the accretion rate in Section \ref{sec:analytics}, and describe our numerical setup in Section \ref{sec:setup}. We analyze our results in Sections \ref{sec:morphology} and \ref{sec:accretionrate} including the accretion rate, drag forces and location of the sonic point. We discuss these results and conclude in Section \ref{sec:discussion}.

\section{Review of Hydrodynamic Accretion} \label{sec:analytics}

\subsection{Spherical Accretion}

Though we consider axisymmetric flow in this work, the streamlines deflect toward the radial direction due to gravity near $m$. As the streamlines converge around the accretor, transverse pressure gradients enforce radial flow. We first assume that the flow is one-dimensional in the radial direction and that it has reached a steady state. For subsonic velocities, there are no shocks and the flow is isentropic.

We now review the theory of spherical hydrodynamic accretion and direct interested readers to chapter 14 of \citet{1983bhwd.book.....S} and chapter 13 of \citet{1971reas.book.....Z} for  in-depth analysis. Consider a fluid with density $\rho$, fluid velocity $\vel$, Mach number $\mach$, and pressure $P$ flowing through a pipe with cross-sectional area $A$. In the classical theory of nozzles and diffusers (with body forces absent), a relation between the cross-sectional area and fluid velocity can be obtained from conservation of mass and momentum:
\be
\frac{dA}{A} = \frac{dv}{v}(\mach^{2}-1). 
\ee
This area-velocity relation demonstrates that the sign of the acceleration of the flow depends both on the sign of $dA$ and whether the flow is subsonic or supersonic.
If a gravitational potential is included in the conservation of momentum, a modified version of the area-velocity relation is obtained:
\be
\frac{dA}{A} - \frac{d\Phi}{\csq} = \frac{dv}{v}(\mach^{2}-1). \label{eq:velarea}
\ee
Here the sign of $dv$ depends not only on $dA$ but on $dA$ relative to $d\Phi$, meaning that the flow can potentially be accelerated to supersonic velocities upstream of the throat due to gravity. Spherical accretion onto a sink region can be viewed in the same light as a nozzle in which the area is given simply by $A=4\pi r^{2}$ and the throat has area $A_{\rm sink}=4\pi r_{\rm sink}^{2}$, with a potential $\Phi=-Gm/r$. Then $dA=8\pi r dr$ and $d\Phi=Gmdr/r^{2}$, so Equation (\ref{eq:velarea}) becomes
\be
\frac{2}{r} = \frac{\nabla v}{v}(\mach^{2}-1) + \frac{Gm}{\csq r^{2}}.
\ee
To find the sonic point, we set $\mach=1$, yielding
\be
\rsonic = \frac{Gm}{2 c^{2}_{\rm sonic}}. \label{eq:rsonic2}
\ee
We determine the sound speed at the sonic line, $c_{\rm sonic}$, from the Bernoulli constant
\be
B = \frac{\gamma}{\gamma-1}\frac{P}{\rho} + \frac{1}{2}v^{2} + \Phi.
\ee
In steady flow, $B$ is constant along each streamline \citep{2000hsf..book.....C}, which allows us to equate $B_{\rm sonic}=B_{\infty}$. Using the Bernoulli constant in conjunction with (\ref{eq:rsonic2}), we arrive at
\be
\rsonic = \frac{Gm}{B_{\infty}}\frac{5-3\gamma}{4(\gamma-1)}. \label{eq:rsonic}
\ee
This is similar to the calculation of \citet{1983bhwd.book.....S} with the caveat that our $B_{\infty}$ contains a kinetic term $v_{\infty}^{2}/2$ as in \citet{1997A&A...320..342F}. For $\gamma=5/3$, this gives $\rsonic=0$, and for $\gamma>5/3$ it gives $\rsonic<0$. Only for $\gamma<5/3$ is $\rsonic$ positive. However, this calculation neglected the finite radius of the accretor. Within $\rsink$, the pressure and density are negligible, which sets a minimum radius for the sonic surface at $r=\rsink$.


For steady radial flow, the mass accretion rate can be expressed as $\mdot = A \rho v = 4\pi r^{2} \rho v$. Because the flow is isentropic, $P=K\rho^{\gamma}$ for some pseudoentropy $K$, and $\csq = \gamma P/\rho = \gamma K \rho^{\gamma-1}$.
Combining this with (\ref{eq:rsonic2}) gives the density at the sonic line 
\be
\rho_{\rm sonic} = \left( \frac{Gm}{2\gamma K \rsonic} \right)^{1/(\gamma-1)}.
\ee
Placing this and the sound speed into $\mdot$, we obtain
\be
\mdot = 4\pi \sqrt{\gamma K} \left( \frac{Gm}{2\gamma K} \right)^{\frac{1}{\gamma-1}+\frac{1}{2}} \rsonic^{\frac{3}{2}-\frac{1}{\gamma-1}}. \label{eq:mdotgeneral}
\ee
For $\gamma=5/3$, the prefactor simplifies to
\be
\mdot = \pi (\gamma K)^{-3/2} (Gm)^{2} \rsonic^{\frac{3}{2}-\frac{1}{\gamma-1}}.
\ee
However, we encounter the issue that $\rsonic=0$ for $\gamma=5/3$. This is alleviated by the fact that
\be
\lim_{\gamma\rightarrow5/3} \rsonic^{\frac{3}{2}-\frac{1}{\gamma-1}} = \lim_{\gamma\rightarrow5/3} \left(\frac{5-3\gamma}{\gamma-1}\right)^{\frac{3}{2}-\frac{1}{\gamma-1}} = 1.
\ee
Finally, this leads us to
\be
\mdot = \pi (\gamma K)^{-3/2} (Gm)^{2}. \label{eq:mdot}
\ee
We see that for the special case of $\gamma=5/3$, $\mdot$ is independent of $\rsonic$ and of the Mach number. This rate is equal to that found by \citet{1952MNRAS.112..195B} for spherically-symmetric accretion from a gas at rest at infinity, \lp{which we refer to hereafter as $\mdot_{\rm B}$.}


\subsection{Axisymmetric Accretion Rates}

In addition to the accretion rate assuming locally spherial flow discussed above, there are several theoretical predictions and numerical fits for $\mdot$. These include:
\begin{enumerate}[(i)]

    \item The oft-quoted formulation
    \be
    \mdot_{\rm BH} = \frac{4\pi (Gm)^2 \rho_{\infty}}{(c_{s,\infty}^2+v_{\infty}^2)^{3/2}}, \label{eq:bondirate}
    \ee
    proposed by \citet{1952MNRAS.112..195B} as an interpolation between the accretion rates for static and supersonic accretors. The original formula in \citet{1952MNRAS.112..195B} was smaller than (\ref{eq:bondirate}) by a factor of 2, but this was corrected in subsequent works so that in the limit $v_{\infty}\gg c_{s,\infty}$ it agreed with the \citet{1939PCPSHoyleLyttleton} result $\mdot=4\pi(Gm)^{2}\rho_{\infty}/v_{\infty}^{3}$ \citep[see, for example,][]{1985MNRAS.217..367S,ruffert2}.


    \item A set of fitting formulae proposed by \citet{ruffert3} based on their numerical results (their Equations 2 -- 8).
    These formulae were fit to simulations performed at $\mach=0.6$, 1.4 and 10. Interestingly, though only one Mach number below unity was studied, this fit predicts (like Equation \ref{eq:mdot}) that $\mdot$ is independent of Mach number for subsonic accretors.

    \item The analytical interpolation formulae for axisymmetric flows with $\gamma>9/7$ proposed by \citet{1997A&A...320..342F} (their Equations 107 and 108):
    \be
    \mdot_{\rm FR97} = \mdot_{\rm B}\left[1+\frac{\gamma-1}{2}\mach_{\infty}^{2}\right]^{\frac{5-3\gamma}{2(\gamma-1)}} \left[\frac{(\gamma+1)\mach_{e}^{2}}{2+(\gamma-1)\mach_{e}^{2}}\right]^{\frac{\gamma}{\gamma-1}}\left[\frac{\gamma+1}{2\gamma\mach_{e}^{2}-\gamma+1}\right]^{\frac{1}{\gamma-1}},
    \ee
    which describes $\mdot$ in terms of deviations from the spherical Bondi rate $\mdot_{\rm B}$. Here the entropy is characterized by the ``effective Mach number'' $\mach_{e}$, defined as
    \be
    \frac{\mach_{e}}{\mach_{\infty}} = \frac{\sqrt{2/\gamma}}{2^{\gamma}\lambda^{(\gamma-1)/2}} \frac{(\gamma+1)^{(\gamma+1)/2}}{(\gamma-1)^{5(\gamma-1)/4}(5-3\gamma)^{(5-3\gamma)/4}},
    \ee
    with the caveat that $\mach_{e}\geq1$, where $0.5<\lambda<1$ is a parameter which can be tuned such that $\mdot_{\rm FR97}$ approaches the Hoyle-Lyttleton rate for high $\mach_{\infty}$. If the flow is taken to be isentropic ($\mach_{e}=1$), then for $\gamma=5/3$ we recover $\mdot_{\rm FR97}=\mdot_{\rm B}$.
    
\end{enumerate}

In this work, we aim to compare these predictions with the results of numerical calculations. To that end, we now discuss the hydrodynamics solvers with which we tackle this problem.

\section{Numerical Setup} \label{sec:setup}

\begin{table}
\begin{center}
\caption{Comparison of the numerical algorithms in our simulation codes. Numerical values reported are from our $\rsink=0.002\ra$ runs.}
\begin{tabular}{c c c}
\hline
\hfill  & \athena\footnote{\citet{athena++}, \url{https://www.athena-astro.app/}} & \arepo\footnote{\citet{2010mnras.401..791s}, \url{https://arepo-code.org/}} \\
\hline
\hline
\hfill mesh geometry & spherical polar grid & Voronoi tessellation \\
\hline
\hfill mesh refinement & geometric in $r$ & adaptive based on $r$ \\
\hline
\hfill $V_{\rm min}$ ($\ra^{3}$) & $4.33\times10^{-13}$ & $2.56\times10^{-13}$ \\
\hline
\hfill $V_{\rm max}/V_{\rm min}$ & $4.56\times10^{10}$ & $4.47 \times10^{11}$  \\
\hline
\hfill total \# of cells & $3.2\times10^{5}$ & $7\times10^{5}$ \\
\hline
\hfill domain shape & sphere & cube \\
\hline
\hfill mesh motion & static & moving \\
\hline
\hfill gas self-gravity & no & yes, no \\
\hline
\hfill time integration & 2$^{\rm nd}$-order van Leer & 2$^{\rm nd}$-order van Leer \\
\hline
\hfill softening length & N/A & $10^{-5}$ \\
\hline
\hfill $\Delta t_{\rm min}$ ($\ra/c_{s}$) & $1.21\times10^{-7}$ & $10^{-7}$ \\
\hline
\hfill $\eta_{\rm CFL}$ & 0.3 &  0.4\\
\hline
\hfill spatial reconstruction & 2$^{\rm nd}$-order PLM & 2$^{\rm nd}$-order PLM \\
\hline
\hfill slope limiter & van Leer & Superbee, van Leer \\
\hline
\hfill Riemann solver & HLLC, Roe & HLLC, exact iterative \\
\hline
\hfill outer boundary & ghost cells & injection region \\
\hline
\hfill inner boundary & ghost cells & sink region \\
\hline
\label{tab:codecomparison}
\end{tabular}
\end{center}
\end{table}

We perform calculations using the arbitrary Lagrangian-Eulerian code \arepo\ \citep{2010mnras.401..791s} and the Eulerian code \athena\ \citet{athena++}. A comparison of these numerical solvers and mesh structures is summarized in Table \ref{tab:codecomparison}. Though they are both grid codes, they differ markedly in their handling of the mesh. \arepo\ uses a moving, unstructured mesh, utilizing a Voronoi tessellation \citep{1992stca.book.....O} to draw a grid around an arbitrary distribution of points. These mesh-generating points then move at the local fluid velocity, with corrections to ensure roundness of the cells. Additionally, we impose refinement conditions for the maximum cell size as a function of distance from the accretor. We initialize the \arepo\ grid with 10 sub-grids, each with 100 cells, where the smallest subgrid's size is determined such that there are at least 1000 cells surrounding the accretor. Our \athena\ calculations use a static spherical-polar mesh in which the cell spacing in the polar and azimuthal directions is constant, with $n_{\theta}=40$ and $n_{\phi}=10$. In the radial direction, we use $n_{r}=800$ cells whose radial extent is geometric, with each cell differing in radial size by a factor of 1.01 from its neighbors. 

Both codes handle time integration via a second-order van Leer predictor-corrector scheme, and spatial reconstruction with a second-order piecewise-linear method (PLM). We set the Courant–Friedrichs–Lewy number to $\eta_{\rm CFL}=0.3$ or 0.4. Slope limiters are applied using the van Leer limiter with the corrections described in \citet{2014JCoPh.270..784M} to keep the limiter total-variation diminishing. Additionally, both codes use the Harten–Lax–van Leer contact (HLLC) approximate Riemann solver \citep{1994ShWav...4...25T}.

We have tested the effects of other Riemann solvers and slope limiters such as the Roe solver \citep{1981JCoPh..43..357R} in \athena, the exact iterative solver in \arepo\ and the Superbee limiter \citep{1986AnRFM..18..337R} in \arepo. We also tested the effect of gas self-gravity in \arepo\ with $R_{\rm sink}/\ra = 0.02$ and 0.002 but found no significant change to the accretion rate nor the flow morphology in any of these cases. This is due to our choice of physical parameters leading to a gas mass inside $R_A$ to always be a few orders of magnitude smaller than the Jeans mass, as well as the accretor. Therefore, in the scenarios checked here, the gravity is always dominated only by the accretor. We note that the self-gravity of the gas is expected to be important in cases of high density regime, as in common envelope evolution.

\subsection{Boundary Conditions} \label{sec:BCs}

\subsubsection{Outer Boundary}

Our wind-tunnel setup requires gas of a specified state to inflow from one end of the domain and freely exit the other end. In \athena, in which we use a spherical domain, this is accomplished by separately treating the upstream ($\theta<\pi/2$) and downstream ($\theta>\pi/2$) sections of the outer boundary. The upstream section imposes the specified ambient conditions in ghost cells, while the downstream portion is a diode: gas is allowed to outflow with zero-gradient conditions but is not allowed to enter the domain.

\arepo\ uses a cubical domain, handling inflow via an injection region on the upstream end of the box. Mesh cells are created with the specified fluid state within this injection region of width 3 in our internal units (defined in Section \ref{sec:initconds}), where at each time-step all cells within this region are updated for the properties of the ambient flow. Outflow is handled by deleting cells whose mesh-generating points are about to leave the domain. By default, the length of the simulation domain is chosen as $D=32\ra$ to match \citet{ruffert2}. In \arepo, this refers to the side length of the cubical domain, while in \athena\ it refers to the diameter of the spherical domain.

\subsubsection{Inner Boundary}

In both codes, the inner accretion boundary is imposed by creating a region of radius $\rsink$ in which the density and pressure are much lower than their surroundings. In \athena, the inner boundary of the domain is at $\rsink$. Immediately inside of this boundary is a shell of ghost cells with state $(\rho,\vel,P)=(10^{-2}\rho_{\infty},\Vec{0},10^{-2}P_{\infty})$. We take a different approach in \arepo, instead imposing a ``sink'' region within $\rsink$ with a sink particle at its center. Any cell whose vertex falls within this region is eliminated, and its previous mass and momenta are added to those of the sink particle.


\begin{table}
\begin{center}
\caption{Input parameters and derived quantities for all simulations performed in this work: the accretor radius $\rsink$, the type of inner boundary condition, the specific heat ratio $\gamma$, the incoming Mach number $\mach_{\infty}$, the gravitational parameter $Gm$, the accretion radius $\ra$, the width of the simulation domain $D$, the mass accretion rate $\mdot$, the predicted accretion rate $\mdot_{\rm B}$ from (\ref{eq:mdot}) (or (\ref{eq:mdotgeneral}) for $\gamma=4/3$), the ratio of $\mdot$ to its lower limit $\mdot_{\rm geo}$, and the momentum accretion rate $\facc$. All quantities are expressed in code units, as described in Section \ref{sec:initconds}.}

\athena\\
\begin{tabular}{c c c c c c c c c c}
\hline
\hfill $R_{\rm sink}$ & $\gamma$ & $\mach_{\infty}$ & $Gm$ & $R_{A}$ & $D$ & $\mdot$ & $\mdot_{\rm B}$ & $\mdot/\mdot_{\rm geo}$ & $\facc$ \\
\hline
\hfill 1      & 5/3 & 0.6 & 0.18  & 1        & 32   & 8.434 & 0.102 & 4.47     & 4.57     \\
\hfill 0.1    & 5/3 & 0.6 & 0.18  & 1        & 32   & 0.365 & 0.102 & 19.4     & 0.134    \\
\hfill 0.02   & 5/3 & 0.6 & 0.18  & 1        & 32   & 0.144 & 0.102 & 191      & 0.0235   \\
\hfill 0.002  & 5/3 & 0.6 & 0.18  & 1        & 32   & 0.106 & 0.102 & 14100    & 0.00150  \\
\hfill 0.0002 & 5/3 & 0.6 & 0.18  & 1        & 32   & 0.102 & 0.102 & 1350000  & 0.000400 \\
\hline
\hfill 0.002  & 5/3 & 0   & 0.18  & $\infty$ & 32   & 0.106 & 0.102 & $\infty$ & $-1.23\times10^{-6}$ \\
\hfill 0.002  & 5/3 & 0.1 & 0.18  & 36       & 32   & 0.106 & 0.102 & 84300    &  $2.22\times10^{-4}$ \\
\hfill 0.002  & 5/3 & 0.3 & 0.18  & 4        & 32   & 0.106 & 0.102 & 28100    &  $7.71\times10^{-4}$ \\
\hfill 0.002  & 5/3 & 0.8 & 0.18  & 0.56     & 32   & 0.107 & 0.102 & 10600    &  $2.15\times10^{-3}$ \\
\hfill 0.002  & 5/3 & 0.9 & 0.18  & 0.44     & 32   & 0.107 & 0.102 & 9470     &  $2.32\times10^{-3}$ \\
\hfill 0.002  & 5/3 & 1   & 0.18  & 0.36     & 32   & 0.107 & 0.102 & 8540     &  $2.52\times10^{-3}$ \\
\hline
\hfill 36     & 5/3 & 0.1 & 0.18  & 36       & 1152 & 9477  & 0.102 & 23.3     & 939.6 \\
\hfill 4      & 5/3 & 0.3 & 0.18  & 4        & 128  & 120   & 0.102 & 7.96     & 35.16 \\
\hfill 0.56   & 5/3 & 0.8 & 0.18  & 0.56     & 18   & 3.11  & 0.102 & 3.91     & 2.080 \\
\hfill 0.44   & 5/3 & 0.9 & 0.18  & 0.44     & 14.2 & 2.18  & 0.102 & 3.91     & 1.561 \\
\hfill 0.36   & 5/3 & 1   & 0.18  & 0.36     & 11.5 & 1.69  & 0.102 & 4.14     & 1.245 \\
\hline
\hfill 0.002  & 5/3 & 0.3 & 0.045 & 1        & 32 & $7.32\times10^{-3}$ & $6.36\times10^{-3}$ & 1940 & $3.16\times10^{-4}$ \\
\hfill 0.002 & 5/3 & 0.6 & 0.045 & 0.25      & 32 & $7.41\times10^{-3}$ & $6.36\times10^{-3}$ & 983 & $6.31\times10^{-4}$ \\
\hfill 0.002 & 5/3 & 0.8 & 0.045 & 0.141     & 32 & $7.51\times10^{-3}$ & $6.36\times10^{-3}$ & 747 & $8.41\times10^{-4}$ \\
\hline
\hfill 0.02  & 4/3 & 0.6 & 0.18 & 1   & 32     & 0.306 & 0.314 & 405   & 0.0104 \\
\hfill 0.002 & 4/3 & 0.6 & 0.18 & 1   & 32     & 0.306 & 0.314 & 40600 & $-0.261$ \\
\hline
\label{tab:setups}
\end{tabular}

\arepo\\
\begin{tabular}{c c c c c c c c c c}
\hline
\hfill $R_{\rm sink}$ & $\gamma$ & $\mach_{\infty}$ & $Gm$ & $R_{A}$ & $D$ & $\mdot$ & $\mdot_{\rm B}$ & $\mdot/\mdot_{\rm geo}$ & $\facc$ \\
\hline
\hfill 1      & 5/3 & 0.6  & 0.18 & 1      & 32 & 10.15 & 0.102  & 5.38 & 4.93    \\
\hfill 0.1    & 5/3 & 0.6  & 0.18 & 1      & 32 & 0.372 & 0.102  & 19.7 & 0.3  \\
\hfill 0.02   & 5/3 & 0.6  & 0.18 & 1      & 32 & 0.143 & 0.102  & 182 & 0.07\\
\hfill 0.002   & 5/3 & 0.6  & 0.18 & 1      & 32 & 0.102 & 0.102  & 13500 & 0.002 \\
\hline
\hfill 0.02   & 4/3 & 0.6 & 0.18  & 1      & 32 & 0.287 & 0.314  & 381 & $3.5\times10^{-4}$ \\
\hline
\hfill 36      & 5/3 & 0.1  & 0.18 & 36      & 1152 & 13092 & 0.102 & 32.2 & 1050    \\
\hfill 4      & 5/3 & 0.3  & 0.18 & 4      & 128 & 145 & 0.102 & 9.66 & 35    \\
\hfill 0.56      & 5/3 & 0.8  & 0.18 & 0.56      & 18 & 3.68 & 0.102 & 4.67 & 2.49    \\
\hfill 0.44      & 5/3 & 0.9  & 0.18 & 0.44      & 14.2 & 2.48 & 0.102 & 4.53 & 2.06    \\
\hline
\end{tabular}
\end{center}
\end{table}

\subsection{Physical Parameters} \label{sec:initconds}

Unless otherwise specified, we model the gas as ideal with specific heat ratio $\gamma=5/3$. Following \citet{ruffert2}, we choose our code units such that $\rho_{\infty}=c_{s,\infty}=1$, which necessitates $P_{\infty}=K_{\infty}=1/\gamma$. \citet{ruffert2} also chose the characteristic length scale as $\ra=2Gm/v_{\infty}^2=1$, which sets the (initial) mass of the accretor at $Gm=v_{\infty}^2/2$. We deviate from this somewhat by varying $v_{\infty}$ while holding $Gm$ constant at the fiducial value of $(0.6)^{2}/2=0.18$ to study the dependence of $\mdot$ only on $v_{\infty}$. This means that it is not always possible for the simulation domain to span the entire accretion radius since $\ra\rightarrow\infty$ as $v_{\infty}\rightarrow0$. However, the domain size always far exceeds the Bondi radius $R_{B}=2Gm/c_{s,\infty}^{2}=0.36$. For completeness, we also perform several runs at at different value of the accretor mass $Gm=0.045$. \lp{Additionally, we have checked that our results are converged by increasing the linear resolution of our fiducial simulation by a factor of 4 and confirming that our principal results -- the accretion rate and accretion drag -- are not altered significantly.}

\citet{ruffert3} found that the rates of accretion of both mass $\mdot$ and momentum $\facc$ are dependent on the size of the accretor $\rsink$, \lp{\sout{but}} \lp{and their numerical results suggest} that $\mdot$ converges for sufficiently small $\rsink$ (see their Fig. 11). Indeed, the assumptions of Section \ref{sec:analytics} hold provided there exists some radius within which the flow is radial, spherically symmetric, and steady. Motivated by this, we also vary the accretor size from $\rsink/\ra=1$ down to 0.0002, \lp{serving both to confirm the convergence of $\mdot$ and to determine the value to which it converges. To our knowledge, this is the smallest accretor which has been tested, with \citet{ruffert3} using a minimum accretor size of $\rsink/\ra=0.02$ and \citet{2015ApJMacLeod} using $\rsink/\ra=0.01$.}

Although the analytics of Section \ref{sec:analytics} are applicable to the flow near an accretor with $\rsink\ll\ra$ and $\gamma=5/3$, it is also relevant to consider larger accretors and other specific heat ratios. For this reason, we also conduct suites of simulations with $\rsink=\ra$. With accretors of this size, it is important to properly model the flow in the far field. Thus, for these runs we vary the domain size such that $D=32\ra$ is preserved. We also perform two runs with $\gamma=4/3$, in which the sonic line is predicted by Equation (\ref{eq:rsonic}) to occur at a finite radius.

A summary of all of our simulation parameters can be found in Table \ref{tab:setups}. The fork of \athena\ used to perform the simulations in the paper is located at \url{https://github.com/ljprust/athena/tree/windtunnel}, configured with the problem generator file src/pgen/windtunnel.cpp. The parameter file can be found at inputs/hydro/athinput.windtunnel\textunderscore sink and can be modified according to Table \ref{tab:setups} to reproduce our results.

\section{Flow Morphology} \label{sec:morphology}

\begin{figure}
  \includegraphics[width=0.5\textwidth]{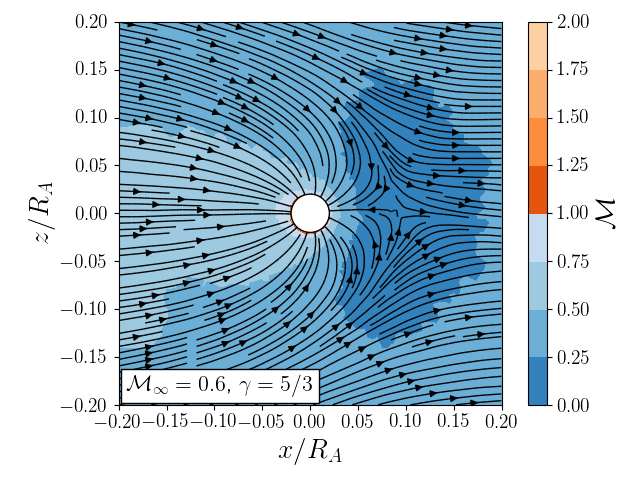}
  \includegraphics[width=0.5\textwidth]{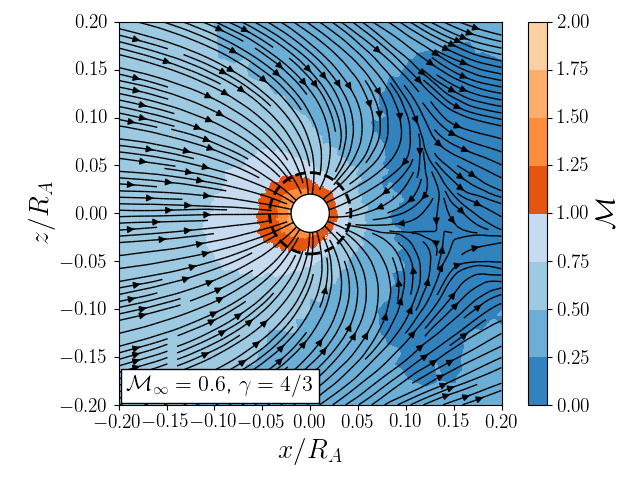} \\
  \includegraphics[width=0.5\textwidth]{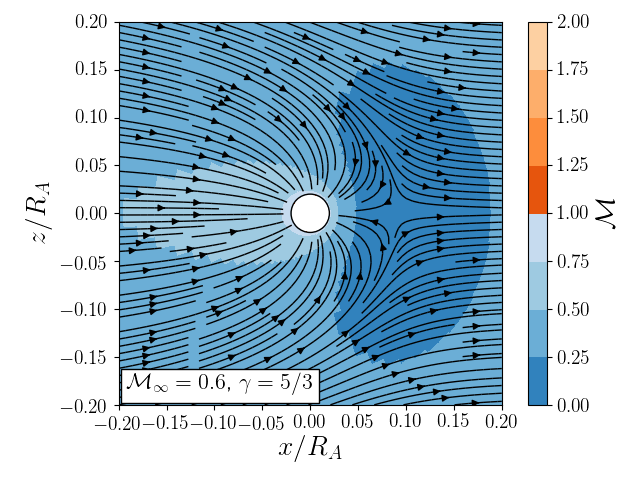}
  \includegraphics[width=0.5\textwidth]{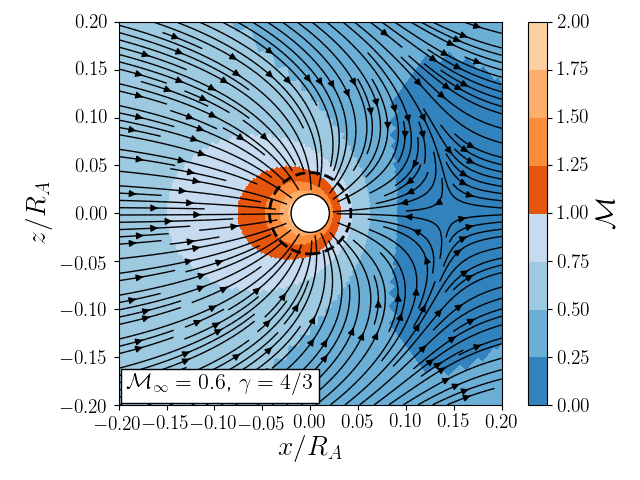}
    \caption{Slice plots of the Mach number in the $\phi=0$ plane in \arepo\ \textit{(top)} and \athena\ \textit{(bottom)} for $\rs=0.02\ra$ with $\gamma=5/3$ \textit{(left)} and $\gamma=4/3$ \textit{(right)}. For $\gamma=4/3$, the dotted line marks the location of the sonic line predicted by (\ref{eq:rsonic}).
    \label{fig:slices}}
\end{figure}

Our simulations eventually settle into a steady state, as defined by the variations in derived quantities. Slice plots of the Mach number in both codes for $\rs=0.02\ra$, $\mach=0.6$ are shown in Fig. \ref{fig:slices} for $\gamma=5/3$ (left panels) and $\gamma=4/3$ (right panels). The overall morphology is very similar between the two codes. The flow accelerates as it nears the accretor, achieving $\mach=1$ at $\rsink$ for $\gamma=5/3$ and at some $r>\rsink$ for $\gamma=4/3$. These features are discussed below, as are the accretion rates and drag forces.


\begin{figure}
\begin{center}
  \includegraphics[width=0.7\textwidth]{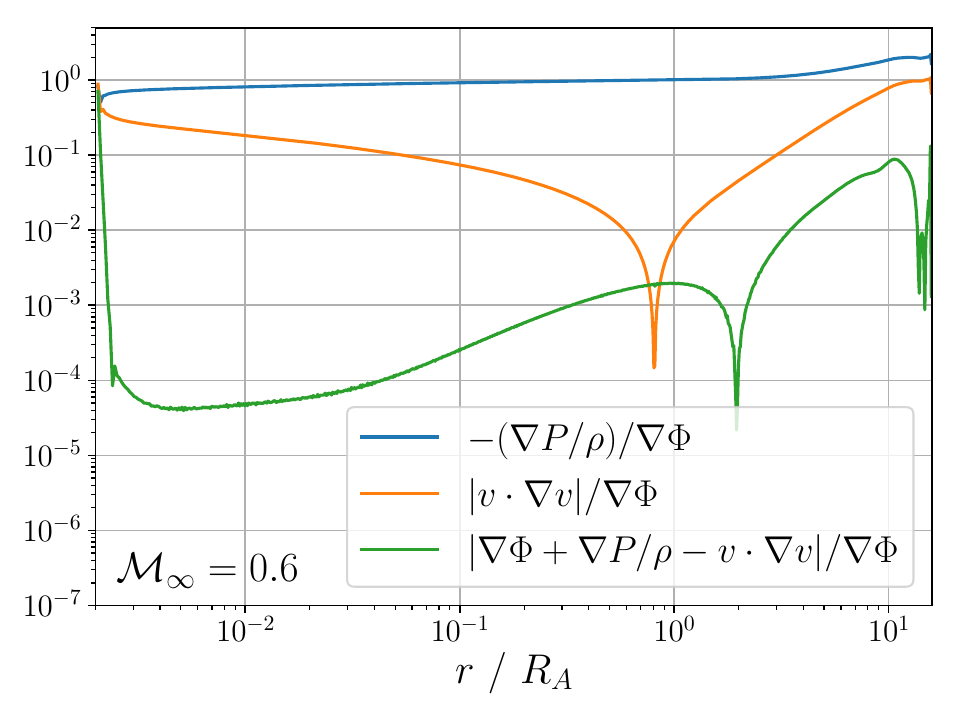}
    \caption{Comparison of the terms in the momentum equation (\ref{eq:momemtum2}) \lp{normalized by the gravitational acceleration} for $\rsink=0.002\ra$, $\mach_{\infty}=0.6$ in \athena. For intermediate radii $10^{-2}\lesssim r/\ra\lesssim 1$, the pressure gradients are comparable to the potential, indicating pseudo-hydrostatic equilibrium.
    \label{fig:hse}}
\end{center}
\end{figure}

\begin{figure}
  \includegraphics[width=0.5\textwidth]{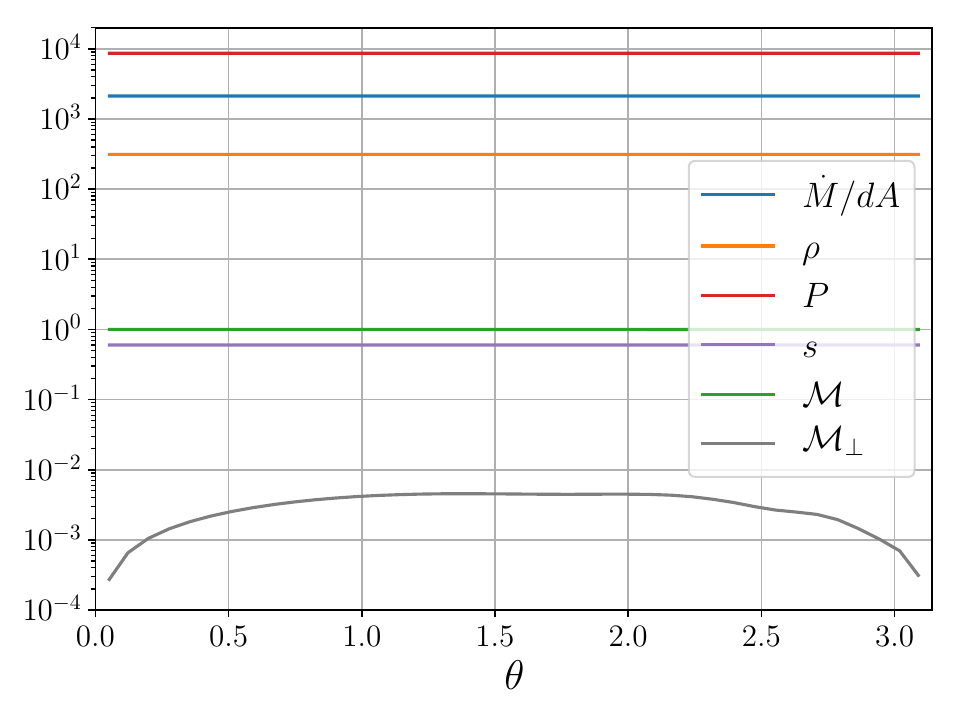}
  \includegraphics[width=0.5\textwidth]{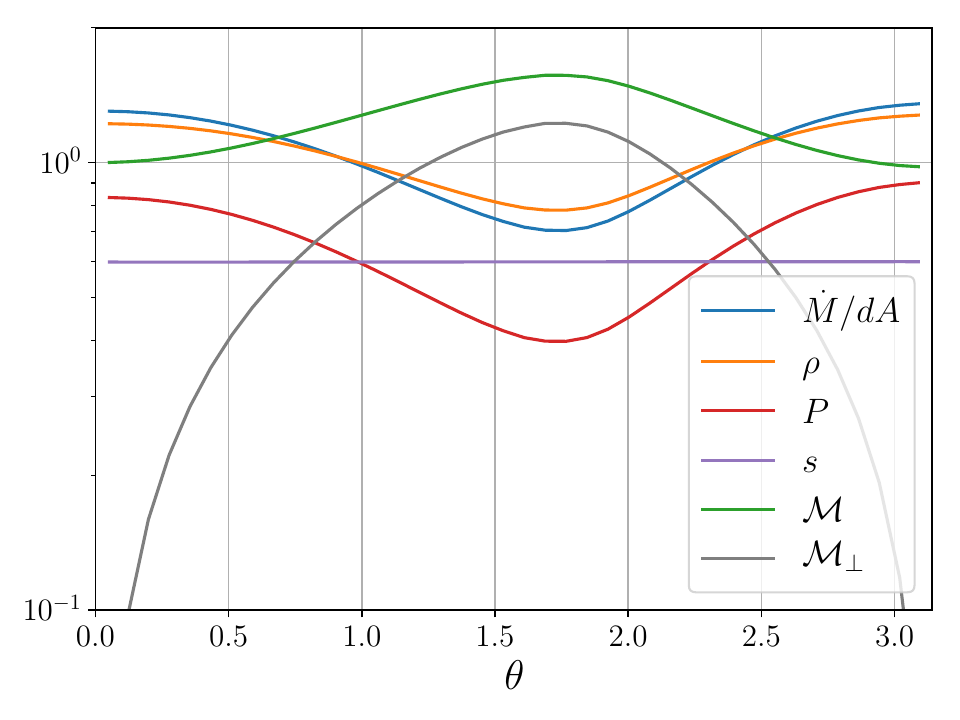}
    \caption{\textit{(Left)} Conditions at the surface of the accretor for $\mach_{\infty}=0.9$ with $\rsink=0.002\ra$ in \athena\ showing that all quantities are uniform. The exception is the Mach number transverse to the radial direction $\mach_{\perp}$, though it becomes much smaller than the radial mach number $\mach\approx 1$ with decreasing accretor size. \textit{(Right)} Surface conditions for $\mach_{\infty}=0.9$ with $\rsink=\ra$. Here the only uniform quantity is entropy, with supersonic flow transverse to the accretor over a portion of its surface. All quantities are presented in code units, as defined in Section \ref{sec:initconds}.
    \label{fig:surface}}
\end{figure}

\begin{figure}
  \includegraphics[width=0.5\textwidth]{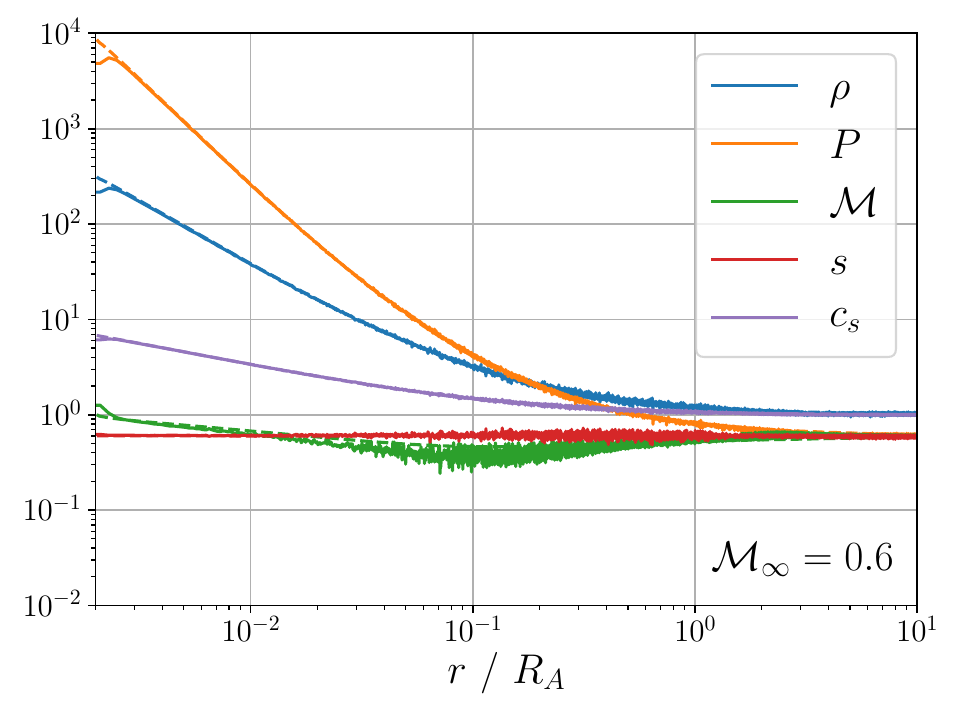}
  \includegraphics[width=0.5\textwidth]{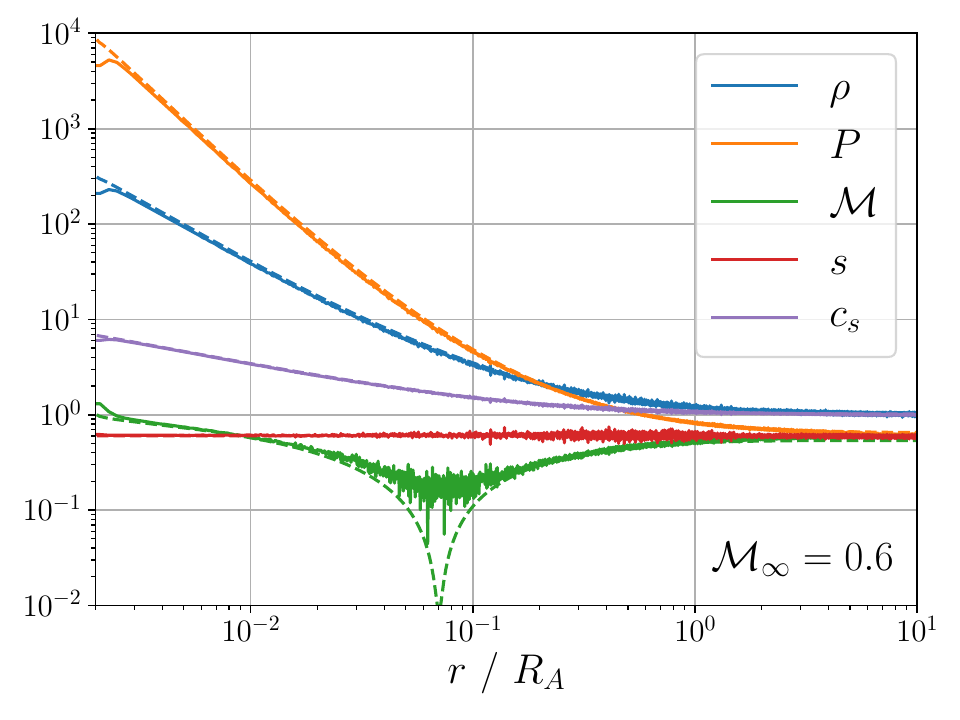}
    \caption{Radial profiles of several quantities for $\rsink=0.002\ra$, $\mach=0.6$ in \arepo\ (solid lines) and \athena\ (dashed lines) along the upstream \textit{(left)} and downstream \textit{(right)} directions.
    \label{fig:profiles}}
\end{figure}

\subsection{Hydrostatic Equilibrium}

The gas is compressed as it approaches the accretor, creating pressure gradients which may be sufficient to enforce a degree of hydrostatic equilibrium. Conservation of momentum can be expressed as
\be
\nabla P/\rho = -\vel\cdot\nabla\vel - \nabla \Phi, \label{eq:momemtum2}
\ee
so the condition for hydrostatic equilibrium ($\nabla P/\rho \approx - \nabla \Phi$) is that the kinetic term is relatively small. We compute each term in Equation (\ref{eq:momemtum2}) along the radial direction and plot them along with their sum in Fig. \ref{fig:hse} for $\rsink/\ra=0.002$, $\mach=0.6$. We see that $\nabla P/\rho\approx-\nabla\Phi$ for $r\lesssim\ra$ with the exception of small radii $r\lesssim10^{-2}\ra$, where pressure support is lost and the flow accelerates. This indicates that much of the flow near the accretor is suspended in near-hydrostatic equilibrium, where pressure gradients in the radial direction are sufficient to prevent gas from free-falling into the accretor. Gradients transverse to the radial direction are also present and may enforce one-dimensional radial flow, as we now discuss.

\subsection{Reaching Spherical Symmetry}


We can learn if the flow near the accretor is radial and spherically-symmetric by checking the flow properties at the surface of the accretor. In Fig. \ref{fig:surface}, we show these properties as well as the Mach number transverse to the radial direction $\mach_{\perp}$ and mass accretion rate per unit area $\mdot/dA$ for the layer of cells adjacent to the sink. We see that for $\rsink=0.002\ra$, $\mach_{\infty}=0.9$ (left panel) all properties are uniform, indicating spherical symmetry has been reached. The exception is the transverse Mach number, though it is small compared to the radial Mach number $\mach=1$ and decreases with accretor size. A large accretor with $\rsink=\ra$, $\mach_{\infty}=0.9$ exhibits a different morphology; spherical symmetry is not obeyed, and both the radial and transverse flow is supersonic over a portion of the surface. We find violation of spherical symmetry for $\mach_{\infty}\gtrsim0.6$ only in the case of $\rsink=\ra$, whereas symmetry is obeyed for $\rsink\leq0.02\ra$ regardless of Mach number.

We compare the radial profiles of several quantities for $\rsink/\ra=0.002$ in Fig. \ref{fig:profiles} along the upstream (left panel) and downstream (right panel) directions in both \arepo\ and \athena. The profiles are very similar between the two codes with the exception of the region very near to the accretor surface, which is likely due to the different methods of implementing the inner boundary. The stagnation point is evident in the downstream Mach number profiles at $r\approx 0.07\ra$. The Mach number reaches unity near $\rsink$ from both sides, as predicted in Section \ref{sec:analytics}, while the entropy remains constant at its initial value $K_{\infty}=1/\gamma$.

\subsection{Sonic Line}

As discussed in Section \ref{sec:analytics}, as gas flows toward the accretor there must exist a point at which it reaches sonic velocity, which for $\gamma=5/3$ occurs at the surface of the accretor $r_{\rm sonic}=\rsink$. We find this to be the case in our simulations, as shown in Fig. \ref{fig:profiles}. For $\gamma<5/3$, the sonic line occurs at a finite radius outside of $\rsink$ as estimated from (\ref{eq:rsonic}). Specifically, for $\gamma=4/3$ we obtain $r_{\rm sonic}=0.0425\ra$, which is overplotted in the lower right panel of Fig. \ref{fig:slices}. We see that the predicted and actual sonic surfaces roughly agree, though the actual sonic surface is not spherically symmetric. Because the region within the sonic surface is causally disconnected from the outside, the accretion rate cannot be affected by the flow within the sonic surface. This means that, unlike $\gamma=5/3$, the accretion rate is unaffected by the size of the accretor provided $\rsink\lesssim r_{\rm sonic}$. Indeed, for $\gamma=4/3$ we find no dependence of $\mdot$ on $\rsink$ (see Table \ref{tab:setups}).

\subsection{Stagnation Point}

Downstream of the accretor, there exists a neutral point in the flow where $\mach=0$. The position of this stagnation point cannot be determined analytically, but an upper bound for its distance $R_{\rm stag}$ from the accretor can be estimated. \citet{1972MNRAS.160..255L} argued that $R_{\rm stag}$ may be of order the accretion radius, as it separates streamlines which will escape to infinity from those which will intersect the accretor. From an energy argument, it can then be estimated that $Gm/R_{\rm stag}\sim v_{\infty}^{2}/2\rightarrow R_{\rm stag}\sim 2Gm/v_{\infty}^{2}$. However, \citet{1972MNRAS.160..255L} also claimed that this likely sets an upper bound for $R_{\rm stag}$ since collisions in the flow will lessen the kinetic energy. The zero-crossing of the radial velocity is sharply-defined in our \athena\ results, allowing us to measure the position of the stagnation point. The results are shown in the right panel of Fig. \ref{fig:vsrsink}. This substantiates the claim that $\ra$ sets an upper bound on $R_{\rm stag}$, as $R_{\rm stag}\gtrsim\ra$ for only our largest accretor size and decreases with $\rsink$.

\section{Accretion Rate} \label{sec:accretionrate}

\begin{figure*}
\begin{center}
  \includegraphics[width=0.8\textwidth]{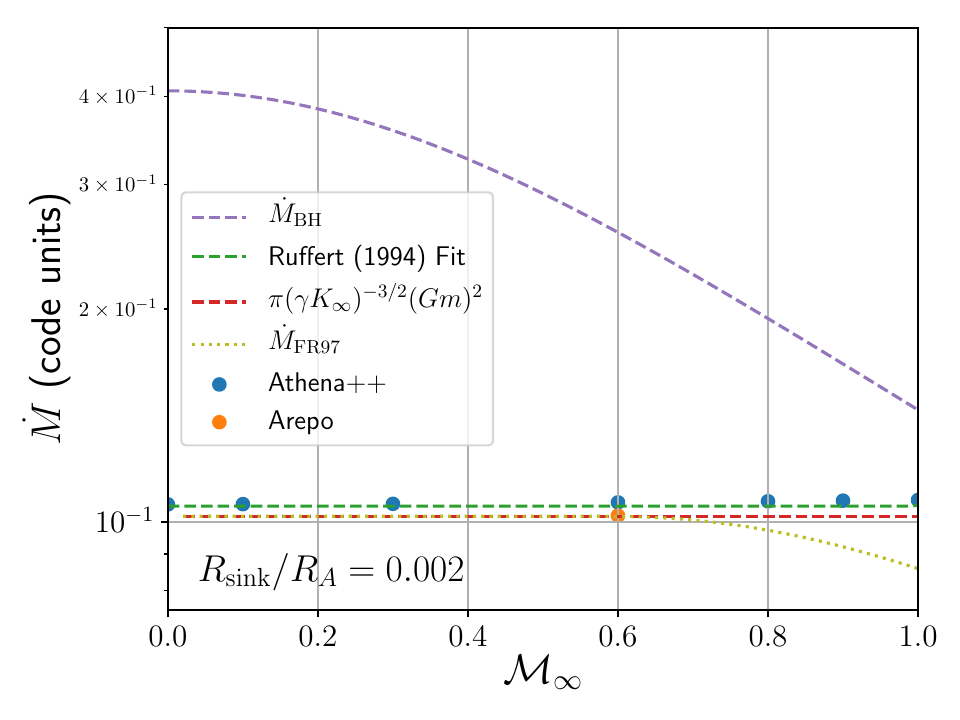}
    \caption{A comparison between the accretion rate obtained from our numerical simulations with $\rsink/\ra=0.002$ and several previous predictions. Our analytics and the fitting formulae of \citet{ruffert3} are consistent with the constant value measured from our simulations, which is lower than the prediction of (\ref{eq:bondirate}). The interpolation of \citet{1997A&A...320..342F} is consistent with our findings over a limited range of $\mach_{\infty}$.
    \label{fig:mdot}}
\end{center}
\end{figure*}

\subsection{Mass Accretion Rate} \label{sec:mdot}

A key objective of this study is the determination of the rate of mass accretion onto the accretor, $\mdot$. We compute this rate in both codes, though through different means. Because \arepo\ takes the mass of all cells which fall within the sink region and add their mass to that of the sink particle, the accretion within one time-step is simply the sum of the masses of these cells. This causes the mass of the accretor to increase by a factor of $\sim1+10^{-5}$ over our integration time. In \athena, it is instead computed by integrating the mass flux over the accretor surface:
\be
\mdot &=& \int_{A} \rho v_{r} dA = \rsink^{2} \sum_{r_{i}=\rsink} \rho_{i} v_{r,i} \Delta\theta_{i} \Delta\phi_{i} \sin\theta_{i}.
\ee

Our numerical results along with these predictions are shown in Fig. \ref{fig:mdot}. We see that the numerical results are consistent with (\ref{eq:mdot}) \lp{\sout{as well as the fit of \citet{ruffert3}}} in that it is independent of Mach number and correctly predict the value of $\mdot$. \lp{This independence has been implicitly thought to exist over at least some range of subsonic $\mach$ but not explicitly verified. The fit of \citet{ruffert3} is also independent of $\mach_{\infty}$ for subsonic flow, though this is not discussed in their work.} Interestingly, the later interpolation by \citet{1997A&A...320..342F} \lp{\sout{is consistent with this value}} \lp{has this property} only for $\mach_{\infty}\lesssim0.7$. The interpolation of the Bondi rate $\mdot_{\rm BH}$ overestimates our measured $\mdot$ for all subsonic velocities. In addition to our fiducial accretor mass $Gm=0.6^{2}/2=0.18$, we also conduct several runs with the smaller value $Gm=0.3^{2}/2=0.045$. Here the expected accretion rate is $\mdot=\pi(\gamma K)^{-3/2}(Gm)^{2}=\pi(0.045)^{2}=6.36\times10^{-3}$. As shown in Table \ref{tab:setups}, the measured $\mdot$ from these runs only slightly exceed this prediction.

\begin{figure}
  \includegraphics[width=0.5\textwidth]{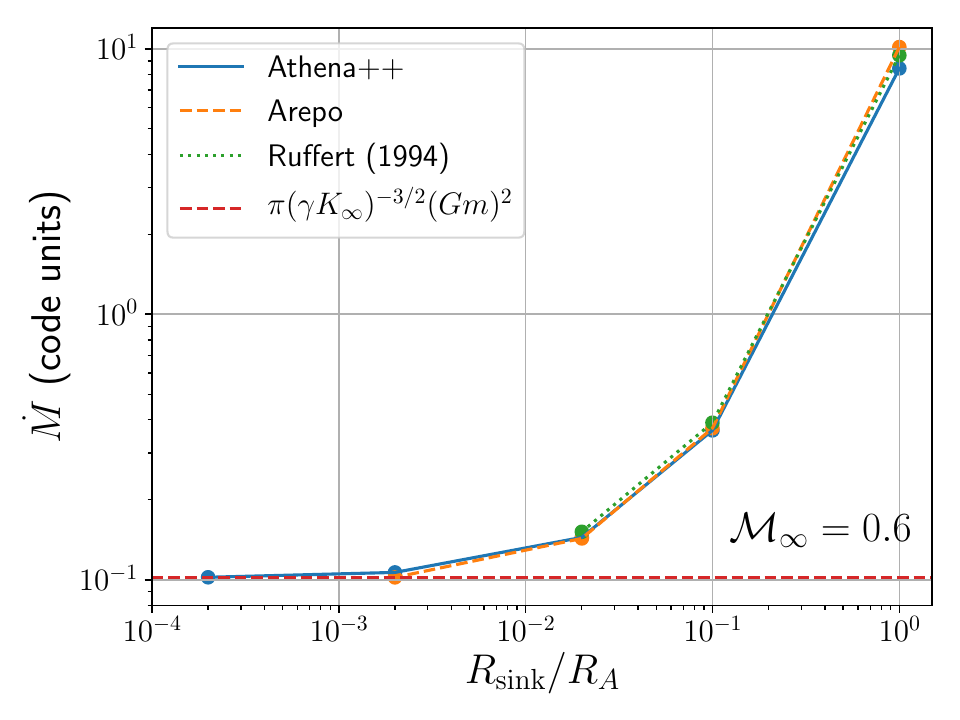}
  \includegraphics[width=0.5\textwidth]{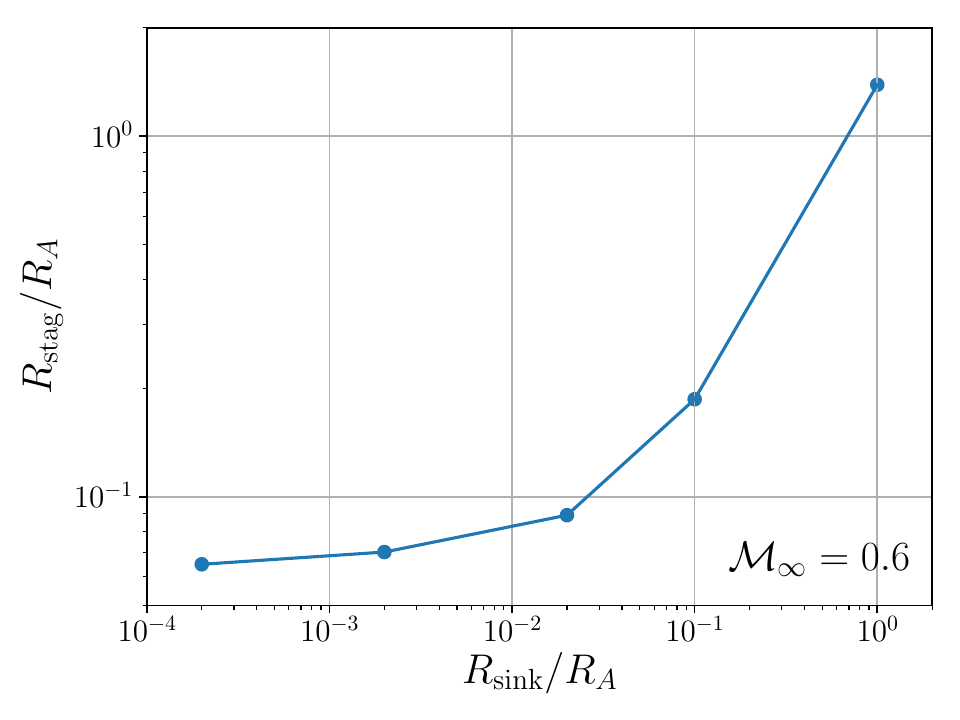}
    \caption{\textit{(Left)} Convergence of $\mdot$ to the predicted value $\pi (\gamma K)^{-3/2} (Gm)^{2}$ given in Equation (\ref{eq:mdot}) with decreasing accretor size $\rsink$. \textit{(Right)} Distance of the stagnation point from the center of the accretor, which falls well below $\ra$ for sufficiently small accretors due to collisions in the flow. For both plots, $\mach_{\infty}=0.6$.
    \label{fig:vsrsink}}
\end{figure}

We also investigate the relation between accretion rate $\mdot$ and accretor size $\rsink$ (left panel of Fig. \ref{fig:vsrsink}), holding $\mach_{\infty}=0.6$ so that $\ra=1$. For larger accretors, our results agree remarkably well with those of \citet{ruffert3}. For smaller accretors, we see that $\mdot$ converges to the predicted value $\mdot_{\rm B}$ in both \arepo\ and \athena.

A lower bound on the accretion rate may be set by the rate at which mass is swept up within a cross-section of $\pi\rsink^{2}$. This bound, which we refer to this as the ``geometric'' accretion rate $\mdot_{\rm geo} = \pi\rsink^{2}\rho_{\infty}v_{\infty}$, should be reached in the limit of a very large accretor $\rsink\gg\ra$. In Table \ref{tab:setups}, we show the ratio of $\mdot$ to $\mdot_{\rm geo}$. We see that $\mdot$ always exceeds this limit by at least a factor of $\approx 4$, coming closest for $\rsink=\ra=0.36$ at high Mach number.

\begin{figure}
    \centering
    \includegraphics[width=0.7\textwidth]{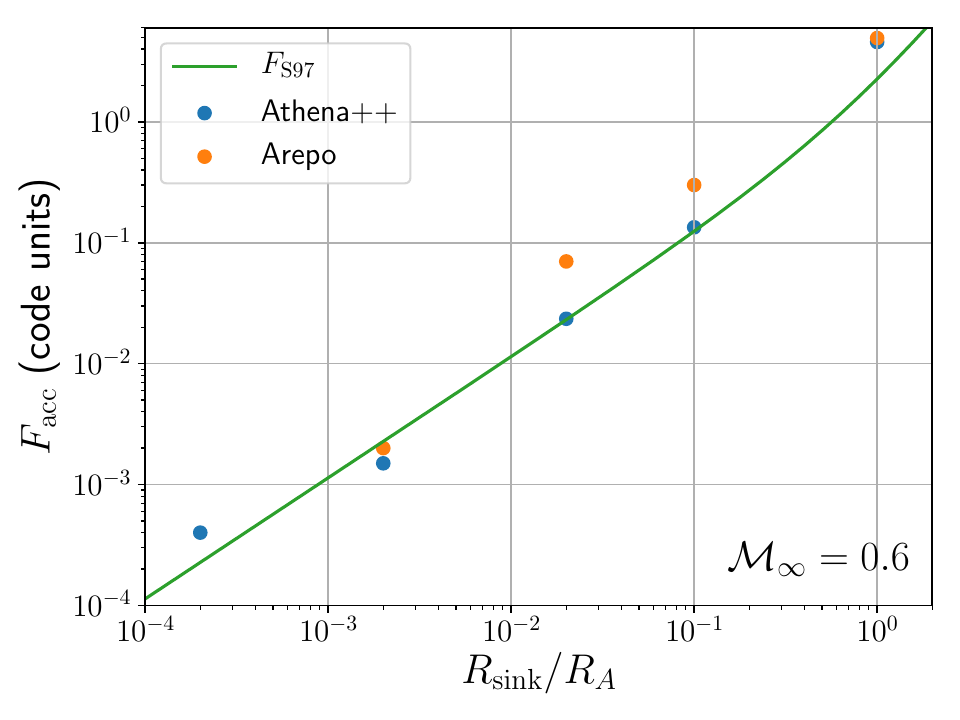}
    \caption{Accretion drag $\facc$ vs accretor size $\rsink$ for $\mach_{\infty}=0.6$, which for small accretors approximately goes as $\facc\propto(\rsink/\ra)$ in agreement with \citet{1972epcf.book.....S}. The dynamical friction ($\fdf\approx 0.1$) is the dominant source of drag for accretors with $\rsink/\ra\lesssim0.1$.}
    \label{fig:facc}
\end{figure}

\subsection{Accretion Drag} \label{sec:drag}

The material passing through the accretor surface carries not just mass, but momentum as well. When the flow is nearly radial, much of this momentum cancels out, though small asymmetries can lead to a net force. In \arepo, the momentum accretion is computed by taking the momentum of all cells which enter the sink region and adding to that of the sink particle. As the sink particle is allowed to move in \arepo, \lp{\sout{this results in}} \lp{it acquires a} small but nonzero velocity \lp{$\approx 2\times 10^{-8}$} \lp{\sout{of the sink}} \lp{in the lab frame for $\rsink/\ra=0.002$ as a result of accreting an amount of gas equal to $\approx 10^{-5}$ of its mass}. In \athena, the static mesh prevents such motion, \lp{though the \arepo\ results demonstrate the validity of this approximation}. Here we compute the rate of momentum accretion along the symmetry axis by again integrating over the surface of the accretor:
\be
\facc &=& \int_{A} \rho v_{r} v_{x} dA = \rsink^{2} \sum_{r_{i}=\rsink} \rho_{i} v_{r,i} v_{x,i} \Delta\theta_{i} \Delta\phi_{i} \sin\theta_{i},
\ee
where the results are shown in Table \ref{tab:setups} and in Fig. \ref{fig:facc}. $\facc$ is generally positive (denoting a force in the direction of the freestream flow), with some exceptions. This can be compared with the gravitational drag due to dynamical friction, which for a subsonic accretor takes on the form
\be
\fdf = -\frac{4\pi(Gm)^{2}\rho}{v^{2}}\left[\frac{1}{2}\ln\left(\frac{1+\mach}{1-\mach}\right)-\mach\right]
\ee
\citep{1999ApJ...513..252O}. For our chosen parameters $\rho_{\infty}=c_{s,\infty}=1$, $Gm=0.18$, and $\mach_{\infty}=0.6$, the dynamical friction is $\fdf\approx 0.1$. Thus, we can infer from Fig. \ref{fig:facc} that dynamical friction is the dominant source of drag for small accretors with $\rsink\lesssim10^{-1}\ra$.

Though the accretion drag depends on the accretor size $\rsink$, unlike the mass accretion rate we do not find convergence for a sufficiently small sink region. For a non-gravitating body, one would expect the cross-section for accretion drag to be $\sigma=\pi\rsink^{2}$. On the other hand, \citet{1972epcf.book.....S} suggests that for a gravitating accretor the cross-section is given by $\sigma=\pi\rsink^{2}(1+\Theta)$, where $\Theta=\ra/\rsink$ is the Safronov number. For $\Theta\ll 1$, this produces the expected behavior $\sigma=\pi\rsink^{2}$, while in the limit $\Theta\gg 1$ it reduces to $\sigma=\pi\ra^{2}\Theta^{-1}$. Here, we find that the resulting drag
\be
F_{\rm S97} \approx \rho v_{\infty}^{2} \sigma = \rho v_{\infty}^{2} \pi\rsink^{2}(1+\Theta)
\ee
well-approximates the data, as shown in Fig. \ref{fig:facc}. \lp{This is valid only for $\gamma=5/3$, as for other equations of state the cross-section is altered due to the sonic surface.}

\section{Discussion and Conclusions} \label{sec:discussion}

In this paper, we have studied Bondi-Hoyle-Lyttleton accretion using the \athena\ Eulerian code as well as the moving-mesh code \arepo. We assume axisymmetry and isentropic flow, which is relevant to subsonic accretors. These codes give nearly identical results despite their many differences in both the numerical evolution of the fluid and in the measurement of derived quantities, indicating that our results are robust to the numerical scheme used. We show that the gas within the accretion radius is suspended in hydrostatic equilibrium, with the exception of a narrow shell adjacent to the accretor, where the gas accelerates to sonic velocity. \lp{For $\gamma=5/3$, this results in an accretion rate which is independent of Mach number and is equal to the spherical Bondi rate \citep{1952MNRAS.112..195B}, depending only on the accretor mass and the gas pseudoentropy. This result suggests a physical explanation in which the flow near to the accretor surface can be approximated as radial and spherically-symmetric. For subsonic flow, there are no shocks and therefore no change in entropy.} \lp{Previously this Mach number invariance has not been explicity confirmed, resulting in many works still using the interpolation of the Bondi rate $\mdot\propto (v_{\infty}^{2}+c_{s,\infty}^{2})^{3/2}$ for subsonic flow.}

We contrast this result with several previous predictions. We find that the interpolated Bondi rate (\ref{eq:bondirate}) \citep{1952MNRAS.112..195B} overestimates $\mdot$ for $\mach<1$, while our rate is consistent with the interpolation of \citet{1997A&A...320..342F} only for $\mach\lesssim0.7$. Our results are consistent with the fitting formulae of \citet{ruffert3} -- which is also devoid of Mach number dependence in the subsonic regime -- while offering a much simpler calculation.

For $\gamma>5/3$, the accretion rate (\ref{eq:mdot}) is not valid due to the vanishing of the sonic radius. However, for $\gamma<5/3$ the sonic surface occurs at a finite radius outside of the accretor, given by Equation (\ref{eq:rsonic}). Indeed, the case $\gamma=4/3$ is of interest when the gaseous medium is radiation-pressure dominated. Our runs with $\gamma=4/3$ are consistent with theory: the actual location of the sonic surface agrees reasonably well with (\ref{eq:rsonic}), and the $\mdot$ corresponding to this sonic radius agrees with the rate measured from our simulations to within several percent. Additionally, unlike the case of $\gamma=5/3$, the accretion rate is independent of accretor size provided the accretor is enclosed by the sonic surface.

This rate is applicable to accretors without hydrodynamically relevant surfaces (e.g. black holes) moving through fairly homogeneous gaseous media. This occurs, for example, when black holes accrete from the interstellar medium, or in binary stellar processes such as common envelope evolution. In particular, if the accretor is engulfed by a star, both the convective and radiative regions of stellar structure may be treated by the formalism in Section \ref{sec:analytics}. Fortunately, the $\gamma$ values that cannot make use of this formalism, $\gamma>5/3$, are rarely relevant for accretion flows -- though see \citet{2021MNRAS.502.3003R} for a discussion of Bondi accretion with a stiff equation of state in the context of neutron stars.

We find a net force exerted on the accretor due to the acquisition of momentum from accreted gas. \lp{For $\gamma=5/3$}, this accretion drag is in the direction of the freestream flow and its value may be predicted using the Safronov number. However, analytical estimates show that the force from accretion drag is eclipsed by that of the dynamical friction when $\rsink\ll\ra$.

Simulations of larger accretors with $\rsink=\ra$ exhibit several qualitative differences from small ones. Here $\mdot$ far exceeds the spherical Bondi rate and approaches the lower bound $\mdot_{\rm geo}$, which is the limit in which gravity is ignored and matter is simply swept up by the motion of the accretor. For sufficiently high Mach number, the radial, spherically-symmetric solution is absent, instead exhibiting supersonic flow in both the radial and transverse directions. This indicates that accretion onto objects with radii comparable to their accretion radius cannot be modeled by the formalism of Section \ref{sec:analytics}.

In this paper, we have treated the gas as homogeneous, neglecting gradients which are expected to be present in many astrophysical scenarios. These gradients could create an angular momentum barrier to accretion or violate our assumption of radial flow near the accretor. Turbulent media are not explored here, and may affect the results. We have also not considered supersonic accretors, which would result in the accretion of high-entropy shocked material. According to (\ref{eq:mdot}), this would reduce $\mdot$. We leave the study of supersonic accretors to future work.

\begin{acknowledgments}
This work was performed in part at the Aspen Center for Physics, which is supported by National Science Foundation grant PHY-2210452. This research was supported in part by grants NSF PHY-1748958 and PHY-2309135 to the Kavli Institute for Theoretical Physics (KITP). LJP is supported by a grant from the Simons Foundation (216179, LB). HG acknowledges support for the project from the Council for Higher Education of Israel. HBP acknowledges support for this project from the European Union's Horizon 2020 research and innovation program under grant agreement No 865932-ERC-SNeX. The work of FKR is supported by the Klaus Tschira Foundation. Computational resources for the calculations performed in this work were provided by the Mortimer HPC System, funded by NSF Campus Cyberinfrastructure Award OAC-2126229 and the University of Wisconsin-Milwaukee. We use the Matplotlib \citep{Hunter:2007} and SciPy \citep{2020SciPy-NMeth} software packages for the generation of plots in this paper.
\end{acknowledgments}

\software{\arepo\ \citep{2010mnras.401..791s},
          \athena\ \citep{athena++},
          Matplotlib \citep{Hunter:2007},
          NumPy \citep{harris2020arrayNUMPY},
          SciPy \citep{2020SciPy-NMeth}
          }

\bibliography{references}{}

\begin{thebibliography}{}
\expandafter\ifx\csname natexlab\endcsname\relax\def\natexlab#1{#1}\fi
\providecommand{\url}[1]{\href{#1}{#1}}
\providecommand{\dodoi}[1]{doi:~\href{http://doi.org/#1}{\nolinkurl{#1}}}
\providecommand{\doeprint}[1]{\href{http://ascl.net/#1}{\nolinkurl{http://ascl.net/#1}}}
\providecommand{\doarXiv}[1]{\href{https://arxiv.org/abs/#1}{\nolinkurl{https://arxiv.org/abs/#1}}}

\bibitem[{{Bondi}(1952)}]{1952MNRAS.112..195B}
{Bondi}, H. 1952, \mnras, 112, 195, \dodoi{10.1093/mnras/112.2.195}

\bibitem[{{Chapman}(2000)}]{2000hsf..book.....C}
{Chapman}, C.~J. 2000, {High Speed Flow}

\bibitem[{{De Young} {et~al.}(1980){De Young}, {Condon}, \&
  {Butcher}}]{1980ApJ...242..511D}
{De Young}, D.~S., {Condon}, J.~J., \& {Butcher}, H. 1980, \apj, 242, 511,
  \dodoi{10.1086/158484}

\bibitem[{{Edgar}(2004)}]{2004EdgarReview}
{Edgar}, R. 2004, \nar, 48, 843, \dodoi{10.1016/j.newar.2004.06.001}

\bibitem[{{Foglizzo} \& {Ruffert}(1997)}]{1997A&A...320..342F}
{Foglizzo}, T., \& {Ruffert}, M. 1997, \aap, 320, 342,
  \dodoi{10.48550/arXiv.astro-ph/9604160}

\bibitem[{Harris {et~al.}(2020)Harris, Millman, van~der Walt, Gommers,
  Virtanen, Cournapeau, Wieser, Taylor, Berg, Smith, Kern, Picus, Hoyer, van
  Kerkwijk, Brett, Haldane, del R{\'{i}}o, Wiebe, Peterson,
  G{\'{e}}rard-Marchant, Sheppard, Reddy, Weckesser, Abbasi, Gohlke, \&
  Oliphant}]{harris2020arrayNUMPY}
Harris, C.~R., Millman, K.~J., van~der Walt, S.~J., {et~al.} 2020, Nature, 585,
  357, \dodoi{10.1038/s41586-020-2649-2}

\bibitem[{{Hoyle} \& {Lyttleton}(1939)}]{1939PCPSHoyleLyttleton}
{Hoyle}, F., \& {Lyttleton}, R.~A. 1939, Proceedings of the Cambridge
  Philosophical Society, 35, 405, \dodoi{10.1017/S0305004100021150}

\bibitem[{{Hunt}(1971)}]{1971MNRAS.154..141H}
{Hunt}, R. 1971, \mnras, 154, 141, \dodoi{10.1093/mnras/154.2.141}

\bibitem[{Hunter(2007)}]{Hunter:2007}
Hunter, J.~D. 2007, Computing in Science \& Engineering, 9, 90,
  \dodoi{10.1109/MCSE.2007.55}

\bibitem[{{Lyttleton}(1972)}]{1972MNRAS.160..255L}
{Lyttleton}, R.~A. 1972, \mnras, 160, 255, \dodoi{10.1093/mnras/160.3.255}

\bibitem[{{MacLeod} \& {Ramirez-Ruiz}(2015)}]{2015ApJMacLeod}
{MacLeod}, M., \& {Ramirez-Ruiz}, E. 2015, \apj, 803, 41,
  \dodoi{10.1088/0004-637X/803/1/41}

\bibitem[{{Mignone}(2014)}]{2014JCoPh.270..784M}
{Mignone}, A. 2014, Journal of Computational Physics, 270, 784,
  \dodoi{10.1016/j.jcp.2014.04.001}

\bibitem[{{Nelson} \& {Benz}(2003)}]{2003ApJ...589..578N}
{Nelson}, A.~F., \& {Benz}, W. 2003, \apj, 589, 578, \dodoi{10.1086/374546}

\bibitem[{{Okabe} {et~al.}(1992){Okabe}, {Boots}, \&
  {Sugihara}}]{1992stca.book.....O}
{Okabe}, A., {Boots}, B., \& {Sugihara}, K. 1992, {Spatial tessellations.
  Concepts and Applications of Voronoi diagrams}

\bibitem[{{Ostriker}(1999)}]{1999ApJ...513..252O}
{Ostriker}, E.~C. 1999, \apj, 513, 252, \dodoi{10.1086/306858}

\bibitem[{{Owen} {et~al.}(2000){Owen}, {Eilek}, \&
  {Kassim}}]{2000ApJ...543..611O}
{Owen}, F.~N., {Eilek}, J.~A., \& {Kassim}, N.~E. 2000, \apj, 543, 611,
  \dodoi{10.1086/317151}

\bibitem[{{Richards} {et~al.}(2021){Richards}, {Baumgarte}, \&
  {Shapiro}}]{2021MNRAS.502.3003R}
{Richards}, C.~B., {Baumgarte}, T.~W., \& {Shapiro}, S.~L. 2021, {Relativistic
  Bondi accretion for stiff equations of state}, Monthly Notices of the Royal
  Astronomical Society, Volume 502, Issue 2, pp.3003-3011,
  \dodoi{10.1093/mnras/stab161}

\bibitem[{{Roe}(1981)}]{1981JCoPh..43..357R}
{Roe}, P.~L. 1981, Journal of Computational Physics, 43, 357,
  \dodoi{10.1016/0021-9991(81)90128-5}

\bibitem[{{Roe}(1986)}]{1986AnRFM..18..337R}
---. 1986, Annual Review of Fluid Mechanics, 18, 337,
  \dodoi{10.1146/annurev.fl.18.010186.002005}

\bibitem[{{Ruffert}(1994)}]{ruffert3}
{Ruffert}, M. 1994, \aaps, 106, 505

\bibitem[{{Ruffert} \& {Arnett}(1994)}]{ruffert2}
{Ruffert}, M., \& {Arnett}, D. 1994, \apj, 427, 351, \dodoi{10.1086/174145}

\bibitem[{{Safronov}(1972)}]{1972epcf.book.....S}
{Safronov}, V.~S. 1972, {Evolution of the protoplanetary cloud and formation of
  the earth and planets.}

\bibitem[{{Shapiro} \& {Teukolsky}(1983)}]{1983bhwd.book.....S}
{Shapiro}, S.~L., \& {Teukolsky}, S.~A. 1983, {Black holes, white dwarfs and
  neutron stars. The physics of compact objects}, \dodoi{10.1002/9783527617661}

\bibitem[{{Shima} {et~al.}(1985){Shima}, {Matsuda}, {Takeda}, \&
  {Sawada}}]{1985MNRAS.217..367S}
{Shima}, E., {Matsuda}, T., {Takeda}, H., \& {Sawada}, K. 1985, \mnras, 217,
  367, \dodoi{10.1093/mnras/217.2.367}

\bibitem[{{Springel}(2010)}]{2010mnras.401..791s}
{Springel}, V. 2010, \mnras, 401, 791, \dodoi{10.1111/j.1365-2966.2009.15715.x}

\bibitem[{Stone {et~al.}(2020)Stone, Tomida, White, \& Felker}]{athena++}
Stone, J.~M., Tomida, K., White, C.~J., \& Felker, K.~G. 2020, The
  Astrophysical Journal Supplement Series, 249, 4,
  \dodoi{10.3847/1538-4365/ab929b}

\bibitem[{{Struck} {et~al.}(2004){Struck}, {Cohanim}, \&
  {Willson}}]{2004MNRAS.347..173S}
{Struck}, C., {Cohanim}, B.~E., \& {Willson}, L.~A. 2004, \mnras, 347, 173,
  \dodoi{10.1111/j.1365-2966.2004.07189.x}

\bibitem[{{Taam} \& {Bodenheimer}(1989)}]{1989ApJ...337..849T}
{Taam}, R.~E., \& {Bodenheimer}, P. 1989, \apj, 337, 849,
  \dodoi{10.1086/167155}

\bibitem[{{Toro} {et~al.}(1994){Toro}, {Spruce}, \&
  {Speares}}]{1994ShWav...4...25T}
{Toro}, E.~F., {Spruce}, M., \& {Speares}, W. 1994, Shock Waves, 4, 25,
  \dodoi{10.1007/BF01414629}

\bibitem[{Virtanen {et~al.}(2020)Virtanen, Gommers, Oliphant, Haberland, Reddy,
  Cournapeau, Burovski, Peterson, Weckesser, Bright, {van der Walt}, Brett,
  Wilson, Millman, Mayorov, Nelson, Jones, Kern, Larson, Carey, Polat, Feng,
  Moore, {VanderPlas}, Laxalde, Perktold, Cimrman, Henriksen, Quintero, Harris,
  Archibald, Ribeiro, Pedregosa, {van Mulbregt}, \& {SciPy 1.0
  Contributors}}]{2020SciPy-NMeth}
Virtanen, P., Gommers, R., Oliphant, T.~E., {et~al.} 2020, Nature Methods, 17,
  261, \dodoi{10.1038/s41592-019-0686-2}

\bibitem[{{Zeldovich} \& {Novikov}(1971)}]{1971reas.book.....Z}
{Zeldovich}, Y.~B., \& {Novikov}, I.~D. 1971, {Relativistic astrophysics.
  Vol.1: Stars and relativity}

\end{thebibliography}
\bibliographystyle{aasjournal}



\end{document}